\documentclass[reprint,amsmath,amssymb,prx]{revtex4-2}

\usepackage{xcolor}
\usepackage{graphicx}% Include figure files
\usepackage{dcolumn}% Align table columns on decimal point
\usepackage{bm}% bold math
\usepackage{hyperref}% add hypertext capabilities
\renewcommand{\eqref}[1]{Eq.\,{$\left(\ref{#1}\right)$}}
\newcommand{\figref}[1]{Fig.\,{$\left(\ref{#1}\right)$}}

\begin{document}
	
	%\title{Levitated Ferromagnetic Torsional Oscillators for High-Precision Magnetometry and Probing Exotic Interactions}	
%	\title{Quantum-Limited Readout Force-Gradient Sensing with Superconducting Ferromagnetic Meissner-Levitated Oscillators for Precision Measurements and Fundamental Physics Tests}
	
%	\title{Standard-Quantum-Limit Microwave Readout of Meissner-Levitated Force Sensors for Casimir and Short-Range Gravity Tests}
	
	\title{Field-Tunable Meissner-Levitated Ferromagnetic Microsphere Sensor for Cryogenic Casimir and Short-Range Gravity Tests}

	\author{Yi-Chong Ren}
	\author{Feng Xu}
	\author{Wijnand Broer}
	\author{Xiao-Jing Chen}
	\author{Fei Xue}
	\email{Email: xfei.xue@hfut.edu.cn}
	\affiliation{School of Physics, Hefei University of Technology, Hefei, Anhui 230601, China}%

	\date{\today}% It is always \today, today,

	\begin{abstract}
		Near-field force measurements at submicron separations can probe Casimir effects and hypothetical short-range interactions, but require cryogenic operation and stable, \textit{in situ} control of separation-dependent backgrounds. We propose a self-calibrating quantum force-gradient sensor in which a ferromagnetic microsphere is Meissner-levitated above a type-I superconducting plane, while a bias magnetic field reproducibly tunes the equilibrium gap for \textit{in situ} separation scans without mechanical approach. The force gradient is encoded as a resonance-frequency shift tracked by a phase-locked loop, and the motion is read out with a SQUID-coupled, flux-tunable microwave resonator that provides adjustable measurement strength without optical heating. Using the input--output formalism, we derive the conditions for reaching the standard quantum limit (SQL) and identify a counterintuitive scaling law: because displacement-to-flux transduction increases with microsphere size, larger microspheres require fewer photons to reach the SQL, enabling a pathway to macroscopic quantum metrology. We quantify the trade-off between suppression of electrostatic patch potentials (via Au coating) and eddy-current dissipation, project force sensitivities of $\sim 10^{-19}\,\mathrm{N\,Hz^{-1/2}}$ at millikelvin temperatures, and outline protocols to extract Casimir pressure and constrain Yukawa-type deviations from Newtonian gravity over $0.1$--$10\,\mu\mathrm{m}$.
	\end{abstract}
		
	\keywords{Quantum sensing; Meissner levitation; SQUID; Flux-tunable microwave resonator; Casimir effect; Non-newton gravity}
	
	\maketitle
	
	\section{Introduction}
	
	Near-field force measurements at submicron and micron-scale separations provide access to Casimir physics and to hypothetical short-range interactions beyond the Standard Model, including Yukawa-type deviations from Newtonian gravity \cite{KapnerPRL2007, MurataCQG2015}. The central difficulty is not simply sensitivity, but \emph{separability}: the sought-after signal grows rapidly at small separation, yet so do dominant separation-dependent backgrounds (Casimir, electrostatic patches, and slow drifts), whose imperfect modeling or subtraction often sets the limiting systematic floor \cite{LambrechtArxiv2011, ChenPRL2016, KlimchitskayaRMP2009}. A second, increasingly relevant limitation emerges once cryogenic operation suppresses mechanical thermal noise: continuous displacement readout becomes constrained by the quantum trade-off between imprecision and measurement backaction, with the optimal balance defining the standard quantum limit (SQL) \cite{ClerkRMP2010, CavesPRD1981, BraginskyKhalili1992}. These two constraints---background separability and quantum backaction---are frequently addressed in isolation. Here we propose a unified principle, \emph{self-calibrating quantum force-gradient spectroscopy}: the sensor is designed so that (i) the dominant near-field backgrounds are calibrated \emph{in situ} by controlled scanning of the equilibrium separation, and (ii) the readout is optimized against an SQL-referenced, input--output noise model that quantitatively sets the required measurement strength.
	
	Our implementation uses a ferromagnetic microsphere Meissner levitated above a type-I superconducting plane, forming a superconducting Ferromagnetic Meissner-Levitated Oscillator (FMLO). In this geometry, an external bias field $B_{\rm ext}$  turns the equilibrium separation $h_0$ into a reproducible control parameter. By sweeping $B_{\rm ext}$, the experiment acquires a stiffness-versus-distance spectrum without moving mechanical stages, enabling long-time averaging under stable cryogenic conditions and mitigating drift systematics that commonly complicate clamped microcantilever platforms \cite{HoferPRL2023, GutierrezPRApp2023}. We operate in a force-\emph{gradient} mode: a static gradient $F'(h_0)$ shifts the mechanical resonance frequency, and in the perturbative regime this mapping is linear, \eqref{S2E04}. Frequency-shift metrology is intrinsically robust against slow gain and offset drifts, making it a natural observable for \emph{in situ} calibration protocols \cite{GiessiblRMP2003, LiuPRA2019}.
	
	To read out the FMLO motion at dilution-refrigerator temperatures without optical heating, we couple the mechanical degree of freedom to a SQUID-based flux transducer and a flux-tunable microwave resonator (FTMR) \cite{SchmidtPRApp2024}. Using an input--output formula description of the SQUID+FTMR readout \cite{RoyPRB2018,KuenstnerPRR2025,CarneyPRL2025}, we derive a compact ``design law'' for force-gradient spectroscopy: the full measurement chain is summarized by an equivalent force-noise power spectral density $S_{FF}$ \eqref{eq:SFF_total}, with explicit quantum imprecision and backaction contributions \eqref{S3E05}. Optimizing over measurement strength yields the SQL condition and the corresponding SQL-referenced force noise \eqref{S3E06}, and directly translates into a required intracavity photon number $\bar n=(G_\ast/G_0)^2$ set by the ratio of the optimal coupling $G_\ast$ to the platform bare coupling $G_0$. In this way, the readout theory does not merely interpret data; it constrains sensor design and operating points.
	
	A main conceptual result of this work is that, for a Meissner-levitated ferromagnetic probe, the SQL requirements can exhibit an \emph{inverted size scaling}. Because the magnetic transduction can make the bare dispersive coupling $G_0(R)$ grow rapidly with sphere radius $R$, the photon number required to reach the SQL, $\bar n_{\rm SQL}(R)$, can \emph{decrease} as the sensor becomes more massive---a ``mass-assisted'' route to quantum-limited readout. This scaling, together with realistic resonator dynamic-range constraints, organizes the parameter space for near-field operation and is summarized in \figref{Fig2} and Sec.~\ref{sec:scaling_sql_photonbudget}.
	
	Near-field separations also force a direct connection between systematic control and mechanical dissipation. A conductive Au coating can suppress electrostatic patch forces and stabilize the surface boundary condition \cite{SpeakePRL2003, GarrettPRR2020}, but it also introduces eddy-current damping that raises the thermal-noise floor and alters the achievable SQL proximity. In our framework, this is not a materials aside: it is a metrological trade-space that jointly constrains background systematics, dissipation, and quantum readout requirements.
	
	The ``self-calibrating'' aspect is made explicit in the application protocols. In the Casimir analysis, the emphasis is the \emph{inference method}: data at large $h_0$ are used to calibrate the slowly varying magnetic background, the associated uncertainty is propagated through the subtraction step, and the residual is then used to extract the Casimir pressure and test power-law behavior over a broad separation range (see \figref{Fig3}). Finally, the same calibrated-residual workflow becomes a capstone application to new physics: assuming that dominant backgrounds are modeled and subtracted at the quoted noise level, the remaining force-gradient spectrum can be fit to a Yukawa form to project constraints on non-Newtonian gravity at interaction ranges $0.1$--$10\,\mu\mathrm{m}$ (see \figref{Fig5}) \cite{KapnerPRL2007, SushkovPRL2011, DeccaPRD2007}.
	
	Beyond Casimir and short-range gravity, the broader motivation is macroscopic quantum metrology with levitated, low-loss sensors. The strong $G_0(R)$ scaling highlights an attractive path toward large-radius, far-field operation where conductive coatings can be reduced or eliminated and dissipation can be extremely low \cite{VinantePRApplied2020}. In that regime, the SQUID+FTMR architecture is naturally compatible with quantum resources (e.g., squeezed microwave drives and backaction-evading measurement) that can surpass coherent-state readout while retaining cryogenic stability\cite{SchmidtPRApp2024, KuenstnerPRR2025,CarneyPRL2025}.
	
	The remainder of the paper is organized as follows. In Sec.~II we introduce the FMLO platform and the force-gradient frequency-shift protocol. In Sec.~\ref{sec:mw_readout} we develop the SQUID+FTMR readout model, derive the equivalent force-noise PSD, and formulate SQL-referenced design criteria, including the radius scaling summarized in \figref{Fig2}. In Sec.~\ref{sec:applications} we present the self-calibration and inference workflows for Casimir-pressure extraction and for projected Yukawa constraints. We conclude with outlooks for quantum-limited and quantum-enhanced operation.

	\section{Experimental Platform and Measurement Protocol}

\begin{figure}[b]
		\centering
		\includegraphics[width=\linewidth]{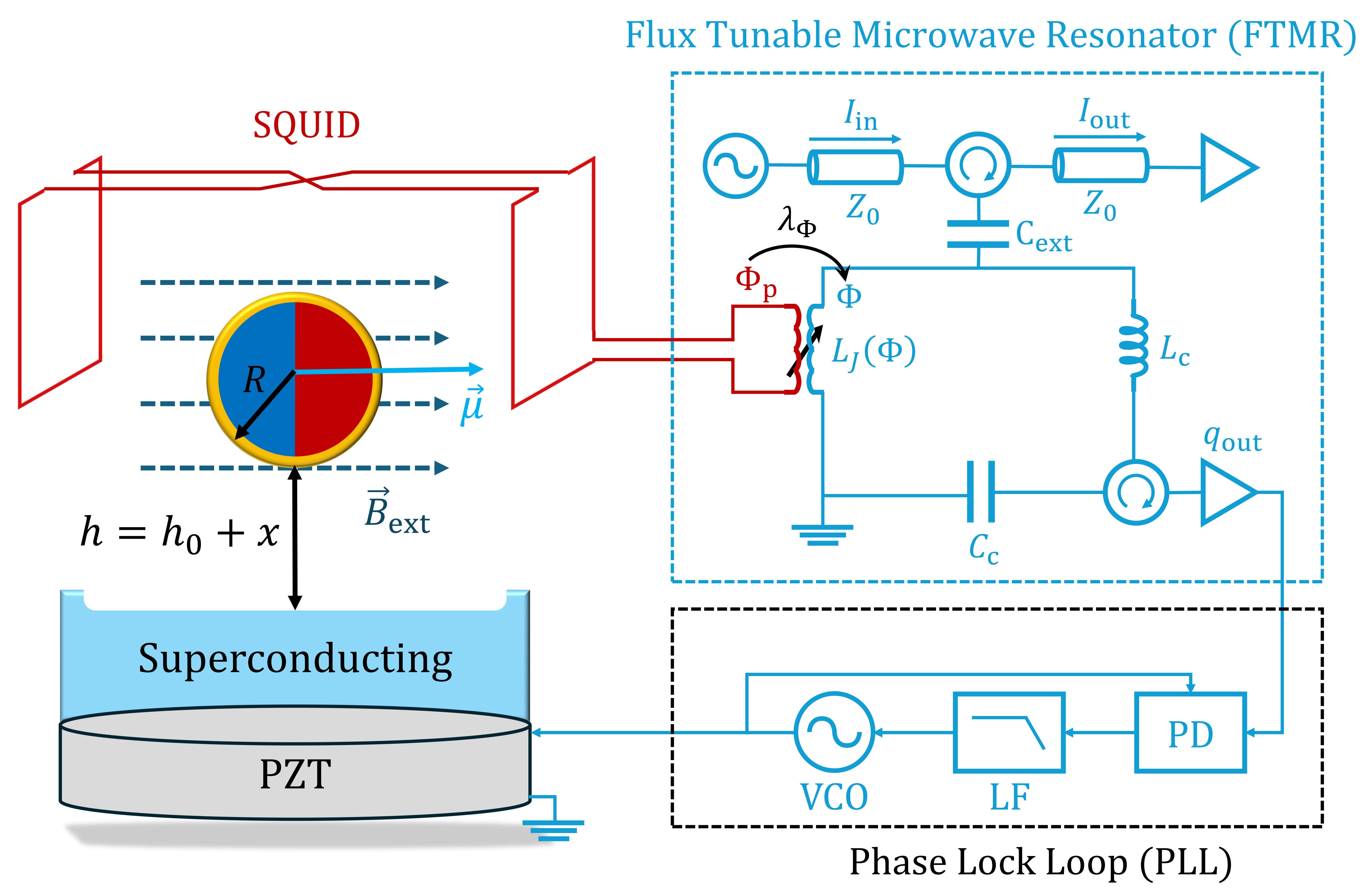}
		\caption{Overview of the proposed force-gradient sensing platform based on a ferromagnetic Meissner-levitated oscillator (FMLO) integrated with a SQUID--flux-tunable microwave-resonator readout and a phase-locked loop (PLL). (a) Mechanical subsystem: a ferromagnetic microsphere (radius $R$) with a conductive Au coating (thickness $d$) is magnetized by a bias field $B_{\mathrm{ext}}$ and Meissner-levitated above a type-I superconducting Pb plane; the equilibrium gap $h_0$ is tuned in situ via $B_{\mathrm{ext}}$, and small vertical motion is denoted by $x$ so that $h=h_0+x$. A piezoelectric actuator (PZT) beneath the plane provides inertial excitation for coherent oscillation. (b) Readout subsystem: the displacement-dependent pickup flux $\Phi_p(x)$ in a SQUID loop is transferred as $\Phi=\lambda_{\Phi}\Phi_p$ to the readout circuit, tuning the Josephson inductance $L_J(\Phi)$ and hence the resonance frequency $\omega_c(\Phi)$ of the flux-tunable microwave resonator (FTMR); the resulting microwave quadrature/phase modulation is detected at the output. (c) PLL: a phase detector (PD) and loop filter (LF) generate an error signal that steers a voltage-controlled oscillator (VCO) to track the mechanical resonance frequency, enabling frequency-shift readout of the force gradient.}
		\label{Fig1}
	\end{figure}

	As shown in \figref{Fig1}, the experimental platform consists of a mechanical subsystem, a readout subsystem, and a PLL subsystem. In the mechanical subsystem, the separation between the microsphere and the plane is described by $ h = h_0 + x $, where $ h_0 $ represents the equilibrium separation and $ x $ is the small displacement around equilibrium. The gold plating on the microsphere provides a stable, chemically inert, and isotropic surface boundary that suppresses electrostatic forces and patch potentials in the near-field~\cite{GarrettPRR2020}. At the same time, the eddy-current losses introduced by the gold layer enhance mechanical damping and increase thermal noise. These effects must be carefully balanced during experiments to select optimal parameters.
	
	The readout subsystem, composed of a SQUID and an FTMR, converts the displacement of the microsphere into measurable current or voltage signals. The PLL subsystem applies a voltage signal through the VCO to the PZT, driving relative motion between the microsphere and the plane. This generates an inertial driving force on the microsphere (within the microsphere's reference frame). Subsequently, the PLL precisely measures the resonant frequency of the mechanical subsystem by comparing the phase of the drive signal with that of the response signal. The entire apparatus is housed in a dilution refrigerator, which provides a millikelvin environment to preserve superconductivity, reduce thermal noise, and suppress thermally induced drifts. To mitigate low-frequency vibration noise and magnetic fluctuations, the system is equipped with multi-stage vibration isolation and multi-layer magnetic shielding.

	Treating the microsphere as a magnetic dipole with moment $ \mu $, the effective potential energy, derived using the image method~\cite{KordyukPRB2006,FuwaPRA2023, HeadleyArXiv2025}, is given by
	\begin{equation}
		U(h) = \frac{\mu_0 \mu^2}{64\pi(R + h)^3} + mg(R + h), \label{S2E01}
	\end{equation}
	where $ \mu_0 $ is the magnetic permeability and $ g $ represents gravitational acceleration. The equilibrium separation $ h_0 $ is determined by minimizing the potential energy, i.e., $ $
	$\left. \partial U/\partial h\right\vert _{h_{0}}=0$, which leads to
	\begin{equation}
		R + h_0 = \left( \frac{3\mu_0 \mu^2}{64\pi mg} \right)^{1/4}. \label{S2E02}
	\end{equation}
	\eqref{S2E02} clarifies the mechanism of \emph{in-situ scanning}: within the region of linear magnetization, $ \mu \simeq \chi V B_{\rm{ext}} $ (where $ \chi $ is the magnetic susceptibility and $ V $ is the volume). Adjusting $ B_{\rm{ext}} $ allows continuous scanning of $ h_0 $ without altering the sample, thus enabling scanning of the force-gradient spectrum as a function of separation.
	
	The system can be approximated as a harmonic oscillator, with stiffness $ k_{\rm{m}} $ and frequency $ \omega_{\rm{m}} $ given by a second-order expansion of the potential energy $ U(h) $ around $ h_0 $:
	\begin{equation}
		k_{\rm{m}} = \left. \frac{\partial^2 U}{\partial h^2} \right|_{h_0} = \frac{4mg}{R + h_0}, \quad \omega_{\rm{m}}^2 = \frac{k_{\rm{m}}}{m} = \frac{4g}{R + h_0}. \label{S2E03}
	\end{equation}
	
	To suppress low-frequency drifts and improve accuracy, we use the force gradient as the primary measurement quantity instead of force. This approach aligns with standard practices in precision measurement fields, such as torsion balance experiments and gravitational wave detectors. When the oscillator is subjected to a static force field $ F(h) $, the effective stiffness and resonant frequency are modified by the force gradient $F'\left(h\right) =\partial F\left( h\right) /\partial h$, with the correction relations $ k_{\rm{eff}} = k_{\rm{m}} - F'(h_0) $ and $ \omega_{\rm{eff}}^2 = k_{\rm{eff}}/m $. In the perturbative regime where $ |F'| \ll k_{\rm{m}} $, the force gradient is linearly related to the frequency shift $ \delta \omega = \omega_{\rm{eff}} - \omega_{\rm{m}} $,
	\begin{equation}
		\frac{\delta \omega }{\omega _{\rm{m}}}=-\frac{F'(h_{0})}{2k_{\rm{m}}}.  \label{S2E04}
	\end{equation}
	Thus, the force gradient can be deduced from the precise measurement of the resonance frequency~\cite{GiessiblRMP2003, LiuPRA2019}. In the experiment, a Phase-Locked Loop (PLL) enables precise locking to the resonance frequency. Combined with the \emph{in-situ scanning} mechanism, this allows for the measurement of the force-gradient spectrum. A statistical analysis is used to subtract the magnetic-gradient background and reduce drift errors~\cite{KohoutekAPL2010, GarrettPRR2020}, yielding the target force gradient $ F'(h_0) $.
	
	Since the experiment focuses on force gradients rather than force itself, establishing a correlation between the equivalent force PSD $ S_{\rm{FF}} $ and force-gradient PSD $ S_{F'F'} $ is crucial. The measurement of the force gradient involves converting a static force gradient $ F' $ into an alternating force $ F(t) = F' A \cos(\omega t) $ by driving the oscillator to produce sinusoidal motion $ A \cos(\omega t) $. Consequently,
	\begin{equation}
		S_{F'F'} \simeq S_{\rm{FF}}/A^{2}. \label{S2E05}
	\end{equation}
	\eqref{S2E05} indicates that increasing the amplitude enhances the magnetic sensitivity of the force gradient. However, the maximum amplitude is constrained by the linear dynamic range of the mechanical system. Assuming the mechanical system permits a relative nonlinear stiffness error $ \varepsilon $ ($ \varepsilon \ll 1 $), the amplitude is constrained by $A\left\vert \partial k_{\rm{m}}/\partial h\right\vert _{h_{0}}\leq\varepsilon k_{\rm{m}}$, yielding $ A \leq \varepsilon(R + h_0)/5 $ for the FMLO.

	\section{Quantum Microwave Readout and Noise Budget}
	
	\label{sec:mw_readout}
	
	In cryogenic Meissner-levitated platforms, thermal force noise can be strongly suppressed, placing the measurement sensitivity in the regime where readout noise is dominant. The readout noise has two quantum-limited components: imprecision (shot) noise, which decreases with measurement strength, and backaction noise, which increases with measurement strength. Their product is constrained by the uncertainty principle, and optimizing over measurement strength yields the standard quantum limit (SQL) \cite{ClerkRMP2010,CavesPRD1981,BraginskyKhalili1992}.
	
	In the near-field configuration of \figref{Fig2}, the thermal noise can nonetheless exceed the SQL. This excess is not intrinsic to the Meissner-levitated system but arises from eddy-current dissipation in the gold coating introduced to mitigate patch-potential systematics\cite{SpeakePRL2003,GarrettPRR2020}, which limits the mechanical quality factor to $Q\sim10^{5}$. In far-field operation, where such a coating is unnecessary, Meissner-levitated ferromagnetic oscillators are expected to reach $Q\sim10^{8}$ \cite{VinantePRApplied2020,FuwaPRA2023}. In the present implementation, a finite microwave photon budget and the FTMR dynamic range constrain the achievable measurement strength, thereby limiting the microsphere size range for which SQL-level readout is attainable.
	
	Compared with optical readout, which can reach extreme precision but may introduce recoil heating and absorption/photothermal channels \cite{JainPRL2016,AspelmeyerRMP2014}, a SQUID+FTMR readout avoids direct optical heating and provides a widely tunable measurement strength: the SQUID enables strong flux-inductance-frequency transduction, while the FTMR supports large intracavity photon numbers to parametrically enhance the effective coupling.
	
	\subsection{SQUID+FTMR model}
	
	\label{sec:squid_ftmr_model}
	
	The FTMR comprises a pickup coil and an LC resonator inductively coupled to a SQUID. A mechanical displacement $x$ modulates the SQUID flux $\Phi$, which tunes the effective Josephson inductance and hence the LC resonance frequency. This frequency modulation is transduced into changes of the resonator electromagnetic quadratures (charge or flux). A coherent drive populates the resonator with a large intracavity photon number $\bar n$, enhancing both the measurement rate and the effective electromechanical coupling.
	
	For a Josephson element, the inductance $L_{J}(\Phi)$ is $\Phi$-periodic. Linearizing about a flux bias point $\Phi_{0}$ and retaining terms to first order in $\delta\Phi=\Phi-\Phi_{0}$, one obtains
	\begin{equation}
		\omega _{\rm{c}}(\Phi )\simeq \omega_{\rm{c}}(\Phi_{0})+\left(d\omega _{\rm{c}}/d\Phi \right)\delta \Phi.  \label{S3E01}
	\end{equation} For small displacements we further approximate $\delta\Phi(x)\simeq(d\Phi/dx)\,x\equiv\eta x$, yielding a dispersive frequency shift $\delta\omega_{\rm{c}}=(d\omega_{\rm{c}}/d\Phi)_{\Phi_{0}}\eta x$.
	
	Treating the electromagnetic mode (FTMR) and the mechanical mode (FMLO) as harmonic oscillators, we introduce bosonic operators $\hat{a}$ and $\hat{b}$ via the canonical pairs $(\hat{\phi},\hat{q})$ and $(\hat{x},\hat{p})$,
	\begin{equation}
		\begin{array}{ll}
			\hat{x}=x_{0}(\hat{b}+\hat{b}^{\dagger }), & \hat{p}=-\left( i\hbar
			/2x_{0}\right) (\hat{b}-\hat{b}^{\dagger }), \\ 
			\hat{\phi}=\phi _{0}(\hat{a}+\hat{a}^{\dagger }), & \hat{q}=-\left( i\hbar
			/2\phi _{0}\right) (\hat{a}-\hat{a}^{\dagger }) ,
		\end{array} \label{S3Eops}
	\end{equation}
	with zero-point amplitudes $x_{0}=\sqrt{\hbar/(2m\omega_{\rm{m}})}$ and $\phi_{0}=\sqrt{\hbar/(2C\omega_{\rm{c}})}$. The Hamiltonian then takes the standard dispersive form
	\begin{equation}
		H_{\rm{sys}}=\hbar\omega_{\rm{c}}\hat{a}^{\dagger}\hat{a}+\hbar\omega_{\rm{m}}%
		\hat{b}^{\dagger}\hat{b}+\hbar G_{0}(\hat{b}+\hat{b}^{\dagger})\hat
		{a}^{\dagger}\hat{a}, \label{S3Ebare}%
	\end{equation}
	where $\omega_{\rm{c}}$ and $\omega_{\rm{m}}$ are the electromagnetic and mechanical resonance frequencies, and $G_{0}=\eta x_{0}\left\vert d\omega_{\rm{c}} /d\Phi\right\vert _{\Phi_{0}}$ is the single-photon (bare) coupling. A drive at frequency $\omega_{\rm{d}}$ displaces the cavity field, $\hat{a}\rightarrow\alpha+\hat{a}$, yielding the linearized Hamiltonian \cite{ClerkRMP2010,AspelmeyerRMP2014,TeufelNature2011}
	\begin{equation}
		H_{\rm{sys}}=\hbar\omega_{\Delta}\hat{a}^{\dagger}\hat{a}+\hbar\omega
		_{\rm{m}}\hat{b}^{\dagger}\hat{b}+\hbar G(\hat{b}^{\dagger}+\hat{b})(\hat
		{a}^{\dagger}+\hat{a}), \label{S3E02}%
	\end{equation}
	where $\omega_{\Delta}=\omega_{\rm{c}}-\omega_{\rm{d}}$ and $G=\sqrt{\bar{n}}\,G_{0}=|\alpha|G_{0}$, where $\bar{n}=|\alpha|^{2}$ is the intracavity photon number.
	
	For resonant driving ($\omega_{\Delta}=0$), the Langevin equations for the mechanical coordinate $(x,p)$ and cavity quadratures $(\phi,q)$ read
	\begin{equation}
		\begin{array}
			[c]{l}
			\dot{x}=p/m,\\
			\dot{p}=-m\omega_{\rm{m}}^{2}x-x_{0}G\phi-\gamma p+F_{\rm{in}},\\
			\dot{\phi}=-(\kappa/2)\phi-\sqrt{\kappa}\,\phi_{\rm{in}},\\
			\dot{q}=-(\kappa/2)q-x_{0}Gx-\sqrt{\kappa}\,q_{\rm{in}},
		\end{array}
		\label{S3E03}
	\end{equation}
	where $\gamma$ and $\kappa$ are the mechanical and electromagnetic damping rates, and $F_{\rm{in}},\phi_{\rm{in}},q_{\rm{in}}$ are the corresponding input forces/quadratures. Using the input--output formula $q_{\rm{out}}(t)=q_{\rm{in}}(t)+\sqrt{\kappa}\,q(t)$, one finds in frequency space
	\begin{align}
		q_{\rm{out}}(\omega)  & =e^{i\theta_{\rm{c}}}q_{\rm{in}}(\omega)+\sqrt{\kappa}\,x_{0}G\,\chi_{\rm{c}}(\omega)\chi_{\rm{m}}(\omega)F_{\rm{in}}(\omega)\nonumber\\
		& -\kappa(x_{0}G)^{2}\chi_{\rm{c}}^{2}(\omega)\chi_{\rm{m}}(\omega)\,\phi_{\rm{in}}(\omega).\label{S3E04}
	\end{align}
	Here $e^{i\theta_{\rm{c}}}=-\chi_{\rm{c}}/\chi_{\rm{c}}^{\ast}=1+\kappa\chi_{\rm{c}}$, $\chi_{\rm{c}}=1/(i\omega_{\rm{c}}-\kappa/2)$ and $\chi_{\rm{m}}=1/[m(\omega_{\rm{m}}^{2}-\omega^{2}-i\gamma\omega)]$ are the susceptibilities of the LC resonator and mechanical resonator respectively.
	
	\subsection{Equivalent force-noise PSD and the SQL}
	
	\label{sec:noise_sql}
	
	Although the inferred observable is a force-gradient--induced frequency shift [\eqref{S2E04}], it is convenient to report sensitivity as an \emph{equivalent} force-noise PSD, obtained from the readout model and readily comparable across sensors; the conversion between force-gradient noise and force noise is given by \eqref{S2E05}. \eqref{S3E04} identifies the three relevant noise channels: $F_{\rm{in}}$ drives the mechanics, $q_{\rm{in}}$ sets imprecision noise, and the conjugate quadrature $\phi_{\rm{in}}$ generates backaction through the coupling.
	
	The total equivalent force-noise PSD is
	\begin{equation}
		S_{\rm{FF}}=S_{\rm{FF}}^{\rm{SN}}+S_{\rm{FF}}^{\rm{BA}}+S_{\rm{FF}}^{\rm{in}},
		\label{eq:SFF_total}%
	\end{equation}
	with quantum contributions
	\begin{equation}
		S_{\rm{FF}}^{\rm{SN}}=\frac{S_{\rm{qq}}^{\rm{in}}}{\kappa\,(x_{0}G)^{2}|\chi
			_{\rm{c}}|^{2}|\chi_{\rm{m}}|^{2}},\  S_{\rm{FF}}^{\rm{BA}}=\kappa(x_{0}G)^{2}
		|\chi_{\rm{c}}|^{2}S_{\rm{\phi\phi}}^{\rm{in}}. \label{S3E05}
	\end{equation}
	For vacuum (or coherent) inputs, $S_{\rm{qq}}^{\rm{in}}=\hbar/(2C\omega_{\rm{c}})$ and $S_{\rm{\phi\phi}}^{\rm{in}}=\hbar C\omega_{\rm{c}}/2$. Increasing $G$ suppresses imprecision while amplifying backaction. The optimal coupling $G_{\ast}$ defined by $S_{\rm{FF}}^{\rm{SN}}(\omega)=S_{\rm{FF}}^{\rm{BA}}(\omega)$ yields SQL-limited readout\cite{ClerkRMP2010,CavesPRD1981,BraginskyKhalili1992},
	\begin{equation}
		G_{\ast}^{2}=\frac{x_{0}^{2}}{2\hbar\kappa|\chi_{\rm{m}}|\,|\chi_{\rm{c}}|^{2}},\ 
		S_{\rm{FF}}^{\rm{SQL}}=\left.S_{\rm{FF}}^{\rm{SN}}+S_{\rm{FF}}^{\rm{BA}%
		}\right\vert _{G_{\ast}}=\frac{\hbar}{|\chi_{\rm{m}}|}. \label{S3E06}%
	\end{equation}
	The corresponding intracavity photon number requirement is $\bar{n}=G_{\ast}^{2}/G_{0}^{2}$.
	
	\subsection{Scaling of the SQL and photon budget}
	
	\label{sec:scaling_sql_photonbudget}
	
	SQL operation is most transparently phrased as a coupling hierarchy: the mechanics sets a target measurement strength $G_{\ast}$, while the platform sets a bare coupling $G_{0}$. A coherent drive increases the effective coupling as $G=\sqrt{\bar{n}}\,G_{0}$, so the photon budget required to reach $G_{\ast}$ depends on the size scaling of $G_{0}$. For a Meissner-levitated ferromagnetic oscillator, $G_{0}$ can grow rapidly with microsphere size because the magnetic moment and geometry enhance the flux response. As a result, larger spheres can reach the SQL at lower $\bar{n}$ even though the mechanical inertia increases. This scaling contrasts with quadrupole-trapped superconducting-microsphere proposals such as SLedDoG \cite{CarneyPRL2025}, where the relevant transduction varies only weakly with microsphere size in the operating regime, typically pushing the photon-number requirement upward with increasing mass for a fixed measurement objective and bandwidth.
	
	In addition to quantum readout noise, the mechanical force noise includes thermal noise $S_{\rm{FF}}^{\rm{T}}$ and drive-induced noise $S_{\rm{FF}}^{\rm{D}}$,
	\begin{equation}
		S_{\rm{FF}}^{\rm{T}}=4k_{B}T\gamma m,\  S_{\rm{FF}}^{\rm{D}}=S_{\rm{VV}}%
		^{\rm{T}}\,d_{33}^{2}k_{\rm{m}}^{2}, \label{eq:mech_force_noise}%
	\end{equation}
	where $k_{B}$ is the Boltzmann constant, and $T$ is the bath temperature. The drive term models additional fluctuations introduced by inertial actuation in PLL measurements, associated with thermal voltage noise across the PZT. Here $S_{\rm{VV}}^{\rm{T}}=4k_{B}TR_{\rm{pzt}}$ is the PZT thermal-voltage PSD for resistance $R_{\rm{pzt}}$, and $d_{33}$ is the piezoelectric coefficient. In the levitated geometry, clamping losses are absent; at high vacuum, gas damping is negligible and eddy-current dissipation in the gold layer dominates, for which we estimate $\gamma=gd\mu_{0}\sigma_{\rm{Au}}/[32(1+h_{0}/R)^{4}]$, where $\sigma_{\rm{Au}}$ is the low-temperature conductivity of gold.
	\begin{figure}[t]
		\centering
		\includegraphics[width=\linewidth]{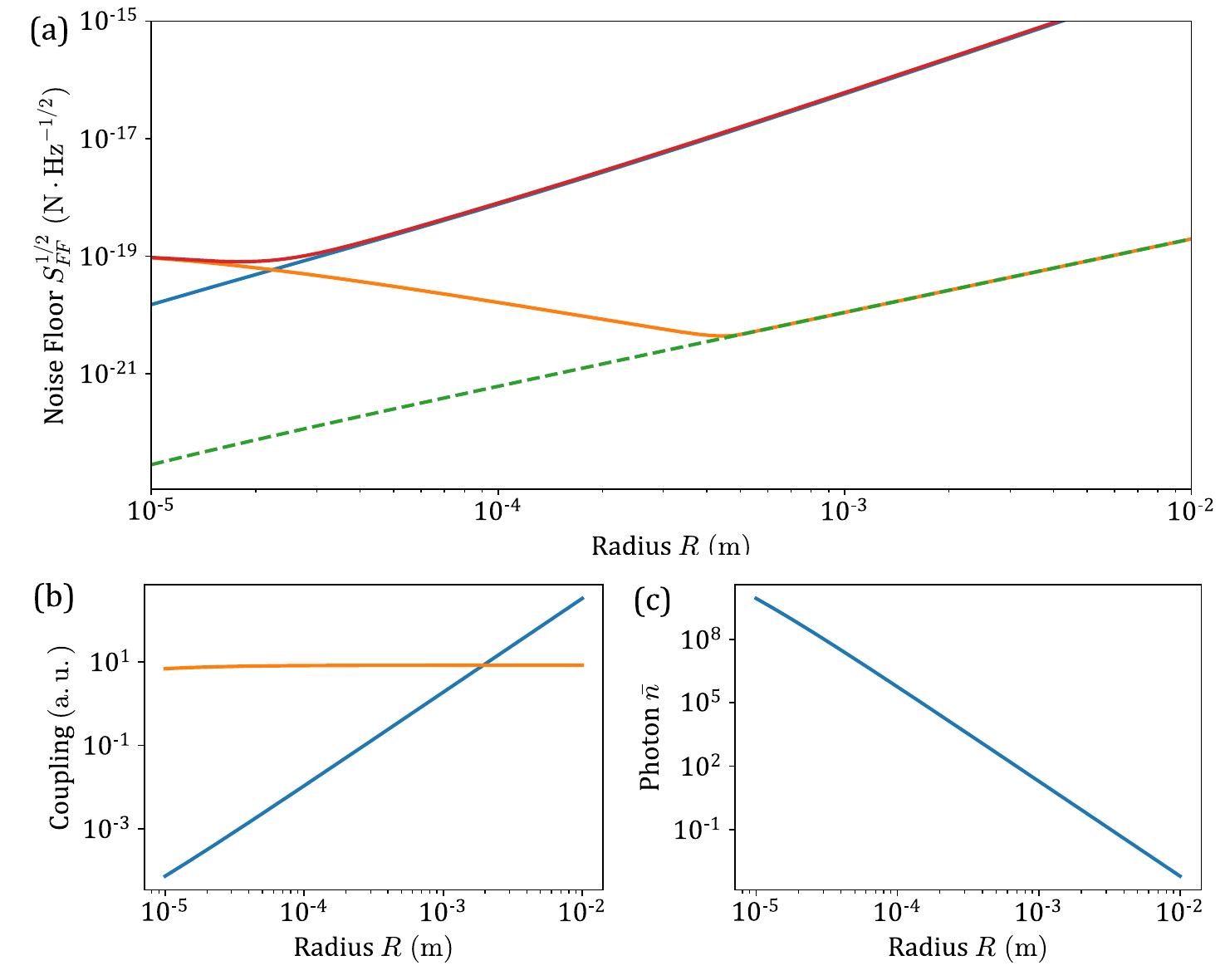} 
		\caption{Radius scaling of the force-noise budget and SQL measurement-strength requirement for near-field FMLO operation (representative parameters: $h_0=1\,\mu\mathrm{m}$, Au thickness $d=100\,\mathrm{nm}$, bath temperature $T=1\,\mathrm{mK}$). (a) Equivalent force-noise amplitude spectral density $\sqrt{S_{FF}}$ evaluated at the mechanical resonance: mechanical noise (solid blue; thermal noise dominated by Au eddy-current damping plus the PZT drive-voltage contribution), quantum readout noise $\bar n\le 400$ (solid orange; achievable with an intracavity photon constraint $\bar n_{\mathrm{SQL}}\le 400$, include imprecision+backaction noises from the SQUID+FTMR readout), the standard quantum limit (SQL, green dashed), and the total noise (solid red; quadratic sum). (b) Bare coupling $G_0$ (blue) and the optimal coupling $G^{\ast}$ required for SQL operation (orange) versus sphere radius. (c) Intracavity photon number $\bar n_{\mathrm{SQL}}=(G^{\ast}/G_0)^2$ required to reach the SQL; under the photon-number cap the readout approaches the SQL only for $R\gtrsim 500\,\mu\mathrm{m}$, whereas for $R\lesssim 30\,\mu\mathrm{m}$ the readout noise exceeds the mechanical noise.}
		\label{Fig2}
	\end{figure}
	
	To quantify the radius dependence, we evaluate the above expressions using representative parameters, the noise contributions and the photon number required to reach the SQL are summarized in \figref{Fig2}.

	\figref{Fig2} shows three trends. (i) Both the mechanical noise and the SQL decrease with decreasing radius, suggesting improved force sensitivity for smaller microspheres. (ii) The increasing magnetic moment of larger  microspheres enhance the displacement-to-flux transduction, thus increase the bare coupling $G_{0}$ rapidly. Therefore, the photon number required by SQL decreases with increasing microsphere radius. (iii) With a photon-number cap ($\bar{n}\leq400$), the quantum readout approaches the SQL only for $R\gtrsim500\,\rm{\mu m}$, while for $R\lesssim30\,\rm{\mu m}$ the readout noise already exceeds the mechanical noise.
	
	For Casimir-force measurements \cite{KlimchitskayaRMP2009,LamoreauxRPP2005}, reducing microsphere size lowers mechanical noise but tightens the photon-budget requirement and increases susceptibility to pull-in. Balancing mechanical and readout noise suggests a radius in the range $10\,\rm{\mu m}\sim 10^{2}\,\rm{\mu m}$, corresponding to a force-noise floor of $10^{-19}\sim 10^{-17}\,\rm{N\cdot Hz^{-1/2}}$, with a minimum practical separation of $\sim 0.1\,\mu\rm{m}$ (smaller separations tend to lead to adhesion). For context, ultrasensitive force measurements have been demonstrated with optically levitated sensors and microfabricated cantilevers. Optical-lattice and feedback-cooled levitated systems report zeptonewton-to-attonewton sensitivities in high vacuum over long integration times \cite{RanjitPRA2016,MonteiroPRA2020,MillenRPP2020}, while cryogenic ultrathin silicon cantilevers achieve force resolutions at the few$\times 10^{-18}\,\rm{N\cdot Hz^{-1/2}}$ level \cite{StoweAPL1997}. These benchmarks provide reference points for the projected FMLO force and force-gradient sensitivities in \figref{Fig2}.

	\section{Applications: Casimir pressure and non-Newtonian gravity}
	
	\label{sec:applications}
	
	The framework developed above enables quantitative projections for near-field Casimir force-gradient measurements and Yukawa-type tests of non-Newtonian gravity \cite{KapnerPRL2007, SushkovPRL2011, DeccaPRD2007}. The FMLO combines low stiffness with low dissipation and is naturally compatible with ultralow temperatures and magnetic shielding. Together with SQUID+FTMR readout, it can maintain high force-gradient sensitivity at submicron and multi-micron separations. In this section, we illustrate the workflow with two examples: a power-law test of the Casimir pressure and a projected constraint on Yukawa non-Newtonian gravity. For clarity, we use simplified toy models; however, this analytical framework can be equally extended to test different dispersive and thermal Casimir models, probe exotic interactions and gravitational waves, or subtract other background forces.
	
	\subsection{Casimir pressure and a power-law test}
	
	\label{sec:casimir_powerlaw}
	
	For ideal parallel conductors, the Casimir pressure follows the characteristic $ h_0^{-4} $ scaling. In the microsphere–plane geometry, the proximity-force approximation gives the Casimir force gradient $ F_{\rm{C}}'(h_0) = \frac{\pi^3 \hbar c R}{120 h_0^4} $, which is related to the equivalent parallel-plate pressure by
	\begin{equation}
		P_{\rm{C}}(h_0) = \frac{F_{\rm{C}}'(h_0)}{2\pi R} = \frac{\pi^2 \hbar c}{240 h_0^4}. \label{S4E01}
	\end{equation}
	
\begin{figure}[h]
		\centering
		\includegraphics[width=\linewidth]{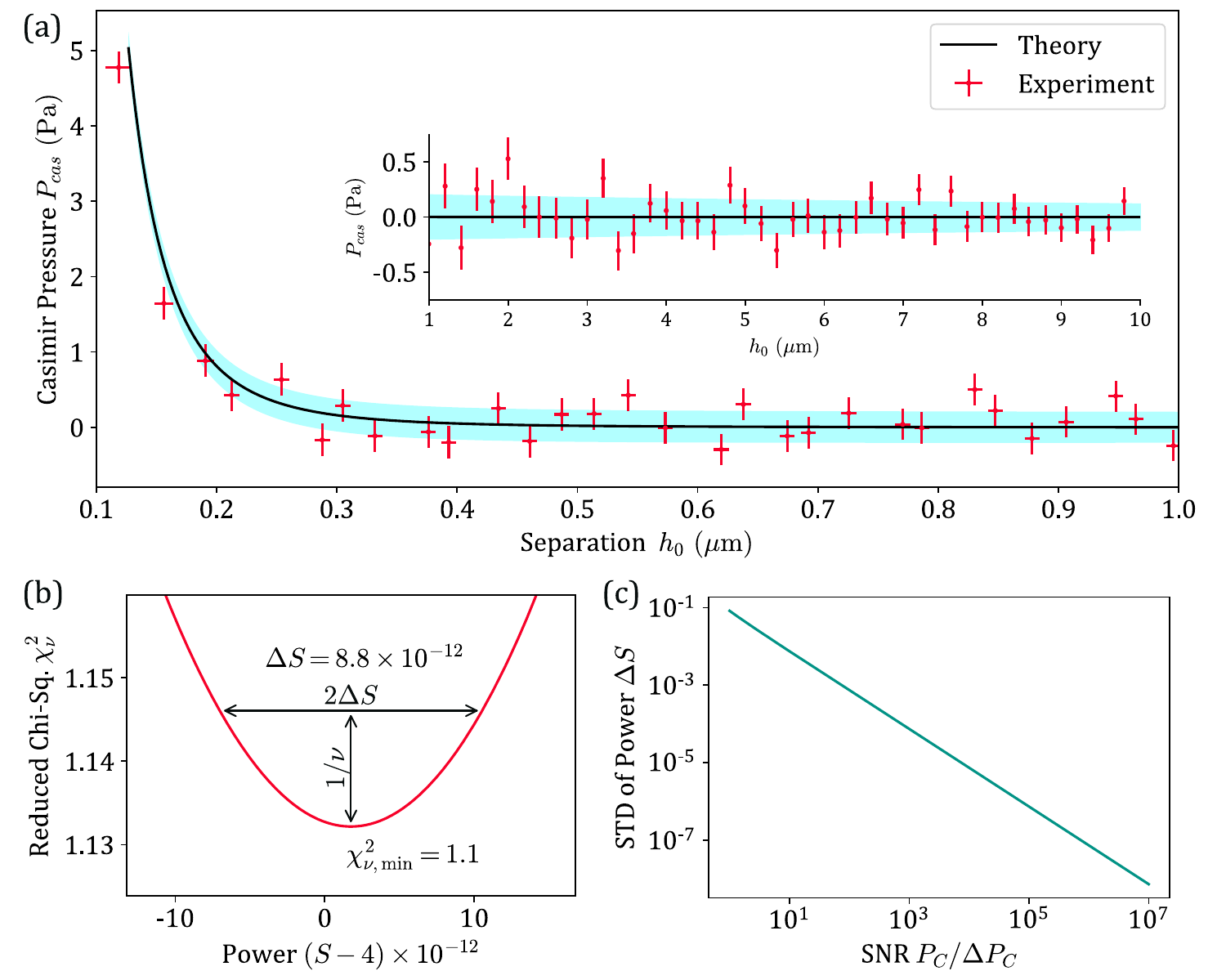}
		\caption{Projected Casimir-pressure extraction and power-law fitting with the FMLO force-gradient protocol. (a) Casimir pressure $P_C(h_0)$ inferred from the force-gradient spectrum after calibrating the slowly varying magnetic background with large-$h_0$ data and subtracting it; symbols show a representative data set generated using the noise budget (integration time $100\,\mathrm{s}$ per separation), and the solid curve shows the nominal Casimir model used for the illustration. The shaded band indicates the propagated $1\sigma$ uncertainty, including the contribution from background calibration/subtraction; for visibility, the error bars and band are multiplied by $10^{9}$ (data values unchanged). Inset: extracted $P_C$ at larger separations ($1$--$10\,\mu\mathrm{m}$), where the Casimir signal is small and the calibration residual is expected to be consistent with zero within uncertainty. (b) Reduced chi-square $\chi_\nu^2$ versus the fitted exponent $S$ in the model $P_C(h_0)=K h_0^{-S}+B$; the best-fit $S_0$ occurs at the minimum of $\chi_\nu^2$, and $\Delta S$ is obtained from the $\Delta\chi^2=1$ criterion. (c) Scaling of the fitted-exponent uncertainty (standard deviation of $\Delta S$) with the per-point signal-to-noise ratio $\mathrm{SNR}\equiv P_C/\Delta P_C$, highlighting that the precision of the exponent test is set primarily by relative (fractional) pressure uncertainty.}
		
		\label{Fig3}
	\end{figure}

	Experimentally, the oscillator is driven at a single frequency, and the resonance is tracked by a PLL, yielding a precision measurement of the effective stiffness (force gradient). Sweeping $ B_{\rm{ext}} $ changes the equilibrium separation $ h_0 $ and produces an \emph{in-situ} stiffness-versus-distance spectrum. The measured gradient contains a slowly varying magnetic contribution and a rapidly decaying Casimir contribution. At sufficiently large $ h_0 $, the Casimir term becomes negligible, so the magnetic gradient can be calibrated by fitting data at $ h_0 \gtrsim 20 \, \mu \rm{m} $ using \eqref{S2E03} and extrapolating the fit to smaller separations. Subtracting this background and converting the residual gradient to pressure yields the spectrum shown in \figref{Fig3}.
	
	To quantify deviations from the ideal power law over a broad separation range, we fit the extracted pressure to
	\begin{equation}
		P_{C}(h_0) = K h_0^{-S} + B,
	\end{equation}
	here $K,B$ and $S$ correspond to the slope, offset and exponent, respectively.
	
	The best-fit parameters $(K, B, S)$ are determined by minimizing the weighted chi-square function,
	\begin{equation}
		\chi^{2}(K, B, S)=\sum_{i=1}^{N}\frac{\left[P_{i}-P_{C}(h_{0,i};K, B, S)\right]^{2}}{(\Delta P_{i})^{2}},
	\end{equation}
	where $P_{i}$ represents the experimentally extracted Casimir pressure at separation $h_{0,i}$, and $\Delta P_{i}$ denotes the corresponding experimental uncertainty (standard deviation). The uncertainties of the fitted parameters (e.g., $\Delta S$) are established via the \textit{$\Delta\chi^{2}$-criterion}. By profiling the remaining parameters---re-optimizing them for each fixed value of $S$---the uncertainty is defined by
	\begin{equation}
		\chi^{2}(S\pm\Delta S)-\chi_{\rm{min}}^{2}=1.
	\end{equation}
	
	To assess the global agreement between the physical model and the experimental data, we utilize the reduced chi-square $\chi_{\nu}^{2}\equiv\chi^{2}/\nu$, where $\nu=N-m$ represents the degrees of freedom for $m$ fitted parameters. As a dimensionless metric, $\chi_{\nu}^{2}$ admits the following physical interpretations: $\chi_{\nu,\rm{min}}^{2} \simeq 1$ indicates that the theoretical model successfully reproduces the observed data within the assigned experimental uncertainties; $\chi_{\nu,\rm{min}}^{2} \gg 1$ suggests a deficiency in the physical model or an underestimation of experimental errors; whereas $\chi_{\nu,\rm{min}}^{2} \ll 1$ implies overestimated errors or the presence of unmodeled physical correlations between data points.

	Importantly, the fitting precision is governed primarily by the \textit{relative} measurement precision rather than the absolute precision. The per-point pressure uncertainty $\Delta P_C$ in \figref{Fig3}(a) includes both the measurement noise and the propagated uncertainty from calibrating the magnetic background using the large-$h_0$ data and subtracting it. Accordingly, we define the per-point SNR as $\rm{SNR}\equiv P_C/\Delta P_C$, i.e., the reciprocal of relative precision $(\Delta P_C/P_C)^{-1}$. As illustrated in \figref{Fig3}(c), improving this relative precision (increasing SNR) directly reduces the fitted-parameter standard deviation and therefore tightens the constraint on the model.
	
	\subsection{Probing non-Newtonian gravity}
	
	\label{sec:yukawa}
	
	A Yukawa-type interaction mediated by a boson of mass $ m_b $ modifies Newtonian gravity between point masses $ m_1 $ and $ m_2 $ separated by $ |r_{12}| $,
	\begin{equation}
		V(|r_{12}|) = -\frac{G m_1 m_2}{|r_{12}|} \left( 1 + \alpha e^{-|r_{12}|/\lambda} \right), \label{S4E02}
	\end{equation}
	where $ \alpha $ is the dimensionless coupling and $ \lambda = \hbar/(m_b c)$ is the interaction range. Precision force experiments constrain $ \alpha(\lambda) $ using, e.g., torsion balances, optomechanics, Casimir platforms, and isoelectronic techniques \cite{DeccaPRL2005, ChenPRL2016, RanjitPRA2016, MonteiroPRA2020}; the FMLO platform is competitive in regimes where force-gradient sensitivity at micron-scale separations is limiting.
	\begin{figure}[b]
		\centering
		\includegraphics[width=\linewidth]{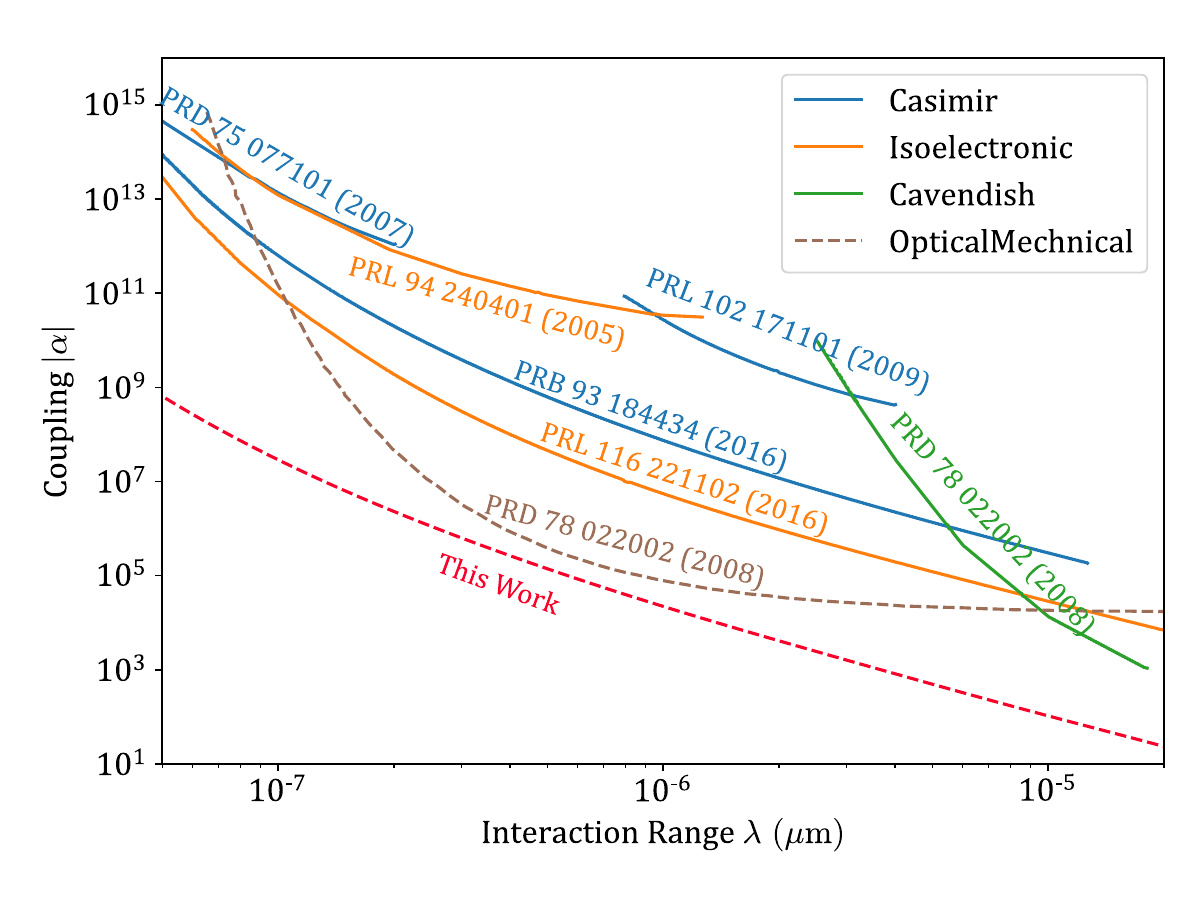}
		\caption{Projected constraints on Yukawa-type deviations from Newtonian gravity in the sphere--plane geometry, expressed as an upper bound on the coupling $|\alpha|$ versus interaction range $\lambda$ [Eq.~(20)]. The red dashed curve (``This work'') shows the projected $1\sigma$ sensitivity obtained by fitting the background-subtracted residual force-gradient spectrum to the Yukawa force-gradient model [Eq.~(21)] using the FMLO noise budget. Colored curves show representative existing laboratory limits from Casimir-force, isoelectronic, Cavendish-type, and optomechanical experiments (labels indicate references and years); parameter values above a given curve are excluded. Over the interaction-length window shown (tick labels correspond to $\lambda\sim 10^{-7}$--$10^{-5}\,\mathrm{m}$, i.e., $0.1$--$10\,\mu\mathrm{m}$), the FMLO projection suggests potential improvements of order $10^{3}$--$10^{6}$ in $|\alpha|$, assuming near-field backgrounds can be modeled and subtracted at the quoted noise level.}

		\label{Fig5}
	\end{figure}
	
	In the geometry shown in \figref{Fig1}, the type-I superconducting plane serves as both the levitation substrate and the source mass. For a plane thickness and lateral size large compared with $ h_0 $ and $ \lambda $, the Yukawa interaction produces a vertical field with force gradient
	\begin{equation}
		F_G'(h_0) = 4\pi^2 \alpha \rho \rho_0 G \lambda^3 e^{-h_0/\lambda} C(R,\lambda), \label{S4E03}
	\end{equation}
	here $ C(R,\lambda) = e^{-2R/\lambda} \left(R/\lambda + 1 \right) + \left( R/\lambda - 1 \right) $, $\rho_0$ and $\rho $ represent the mass densities of the superconducting plane and ferromagnetic microsphere, respectively.
	
	After measuring the total force gradient, subtracting the calibrated magnetic and Casimir contributions yields a residual spectrum. Fitting the residual to \eqref{S4E03} and performing a $ \chi^2 $ analysis provides a projected upper bound (quoted at $ 1\sigma $) on $ |\alpha| $ in \eqref{S4E02}, as shown in \figref{Fig5}.

	Under idealized assumptions, \figref{Fig5} indicates that the superconducting levitation platform could improve bounds on $ \alpha $ by up to several orders of magnitude. In practice, shielding and subtraction of electrostatic and Casimir backgrounds typically require separations above the micron scale, which can reduce sensitivity. A realistic implementation must therefore control additional noise sources and systematics. The present analysis highlights the attainable force-gradient sensitivity and the resulting application reach of FMLO.
	
	\section{Conclusion and Outlook}
	
	In this work we introduced the ferromagnetic Meissner-levitated oscillator (FMLO) as a cryogenic, \emph{in situ}--scannable platform for near-field force-gradient spectroscopy. A key enabling feature is that the equilibrium separation $h_0$ can be tuned reproducibly by the bias field $B_{\mathrm{ext}}$, allowing the sensor to acquire separation-dependent spectra without mechanical approach stages. By operating in the force-\emph{gradient} modality, the interaction signal is encoded as a resonance-frequency shift and can be tracked continuously with a phase-locked loop, providing a drift-robust route to long-integration measurements in the presence of strong separation-dependent backgrounds.
	
	We combined the FMLO with a SQUID-coupled flux-tunable microwave resonator (SQUID+FTMR) to realize a fully cryogenic displacement readout with tunable measurement strength and without optical heating. Using an input--output formula description of the measurement chain, we derived a unified noise budget and expressed it as an equivalent force-noise power spectral density. This SQL-referenced framework makes explicit how imprecision, backaction, thermal noise, and drive noise compete, and it directly connects achievable sensitivity to practical constraints such as resonator dynamic range and the available intracavity photon number. With experimentally motivated parameters, we project force sensitivities of order $10^{-19}\,\mathrm{N\,Hz^{-1/2}}$ at millikelvin temperatures, while quantifying the near-field trade-space in which a conductive Au coating suppresses patch-potential systematics but introduces eddy-current dissipation that raises the thermal-noise floor.
	
	A central conceptual outcome is the inverted scaling of the SQL requirements in the FMLO geometry: because the bare dispersive coupling $G_0$ can grow strongly with sphere radius, the photon number required to approach SQL-level readout can decrease for larger oscillators (Sec.~\ref{sec:scaling_sql_photonbudget}). This ``mass-assisted'' scaling identifies a concrete route toward quantum-limited operation with macroscopic probes. We also established an explicit self-calibration workflow for near-field inference (Sec.~\ref{sec:applications}), in which large-$h_0$ data calibrate the slowly varying magnetic background and the resulting uncertainty is propagated through subtraction before extracting (i) Casimir pressure and power-law behavior and (ii) projected constraints on Yukawa-type deviations from Newtonian gravity over interaction ranges $0.1$--$10\,\mu\mathrm{m}$.
	
	Looking forward, the most direct experimental priorities are improved control of technical backgrounds---magnetic shielding, vibration isolation, and surface preparation/charge management---together with systematic characterization of patch forces and coating-induced dissipation. The radius scaling further motivates a far-field operating regime using large-radius FMLOs, where conductive coatings can be reduced or eliminated, eddy-current losses become negligible, and the exceptionally high mechanical quality factors available in Meissner levitation can be exploited for near-dissipationless inertial and force-gradient sensing. On the quantum-measurement side, the SQUID+FTMR architecture is naturally compatible with quantum-limited amplification and quantum-enhanced protocols, including squeezed microwave drives and backaction-evading measurements, providing a clear pathway from self-calibrating near-field spectroscopy to macroscopic quantum metrology.
	
	\begin{acknowledgments}
		This work was supported by the National Key Research and Development Program of China (Grant No.2024YFF0727902), the Fundamental Research Funds for the Central Universities (Grant No. JZ2025HGTB0176 and No. JZ2024HGTA0182).
	\end{acknowledgments}
	
	\appendix
	
	\section{System Hamiltonian and input--output formalism}
	
	We consider the vertical center-of-mass motion of the ferromagnetic microsphere, described by the displacement $x$, momentum $p$, mass $m$, and mechanical eigenfrequency $\omega_{\mathrm{m}}$. The readout circuit is an $LC$ resonator described by the charge $q$ and flux $\phi$, with total capacitance $C=C_{\mathrm{c}}+C_{\mathrm{ext}}$ and a flux-dependent total inductance
	\begin{equation}
		L(\Phi)=L_{\mathrm{c}}+L_{\mathrm{J}}(\Phi),
	\end{equation}
	where $L_{\mathrm{J}}(\Phi)$ is the Josephson inductance of the SQUID.
	
	The classical Hamiltonian of the coupled system is
	\begin{equation}
		H_{\mathrm{sys}}=\frac{p^{2}}{2m}+\frac{1}{2}m\omega_{\mathrm{m}}^{2}x^{2}
		+\frac{q^{2}}{2C}+\frac{1}{2}C\,\omega_{\mathrm{c}}^{2}(\Phi)\,\phi^{2},
		\label{SA1E01}%
	\end{equation}
	where the cavity resonance frequency is
	\begin{equation}
		\omega_{\mathrm{c}}(\Phi)=\frac{1}{\sqrt{L(\Phi)C}}.
	\end{equation}
	For small flux excursions around a bias point, we linearize $\omega_{\mathrm{c}}(\Phi)$ to first order in $\delta\Phi$ and further linearize the flux induced by a small displacement,
	\begin{equation}
		\delta\Phi(x)\simeq \left(\frac{d\Phi}{dx}\right)x \equiv \eta x,
	\end{equation}
	where $\eta$ is the displacement-to-flux transduction coefficient. Substituting these expansions into \eqref{SA1E01} and keeping terms to leading order yields
	\begin{equation}
		H_{\mathrm{sys}}=\frac{p^{2}}{2m}+\frac{1}{2}m\omega_{\mathrm{m}}^{2}x^{2}
		+\frac{q^{2}}{2C}+\frac{1}{2}C\omega_{\mathrm{c}}^{2}\phi^{2}
		+\frac{1}{2}C\,\omega_{\mathrm{c}}\eta\frac{d\omega_{\mathrm{c}}}{d\Phi}\,x\,\phi^{2}.
		\label{SA1E02}%
	\end{equation}
	
	Upon quantization, we introduce bosonic operators $\hat a$ and $\hat b$ for the electromagnetic and mechanical modes via the canonical pairs $(\hat{\phi},\hat q)$ and $(\hat x,\hat p)$,
	\begin{equation}
		\begin{array}{ll}
			\hat{x}=x_{0}(\hat{b}+\hat{b}^{\dagger }), & \hat{p}=-\left( i\hbar/2x_{0}\right)
			(\hat{b}-\hat{b}^{\dagger }), \\
			\hat{\phi}=\phi _{0}(\hat{a}+\hat{a}^{\dagger }), & \hat{q}=-\left( i\hbar/2\phi _{0}\right)
			(\hat{a}-\hat{a}^{\dagger }),
		\end{array}
		\label{SA1Eops}
	\end{equation}
	with the zero-point amplitudes $x_{0}=\sqrt{\hbar/(2m\omega_{\mathrm{m}})}$ and $\phi_{0}=\sqrt{\hbar/(2C\omega_{\mathrm{c}})}$. The system Hamiltonian becomes
	\begin{align}
		H_{\mathrm{sys}}
		&=\hbar\omega_{\mathrm{m}}\hat{b}^{\dagger}\hat{b}+\hbar\omega_{\mathrm{c}}\hat{a}^{\dagger}\hat{a}
		+\frac{\hbar}{2}G_{0}\left(\hat{b}+\hat{b}^{\dagger}\right)\left(\hat{a}+\hat{a}^{\dagger}\right)^{2},
		\nonumber\\
		&\simeq \hbar\omega_{\mathrm{m}}\hat{b}^{\dagger}\hat{b}+\hbar\omega_{\mathrm{c}}\hat{a}^{\dagger}\hat{a}
		+\hbar G_{0}\hat{a}^{\dagger}\hat{a}\left(\hat{b}+\hat{b}^{\dagger}\right),
		\label{SA1E03b}%
	\end{align}
	where $G_{0}=\eta x_{0}\,(d\omega_{\mathrm{c}}/d\Phi)$ is the bare (single-photon) dispersive coupling rate. In the deriving process of  \eqref{SA1E03b}, we have applied the rotating-wave approximation and dropped the counter-rotating (two-photon) terms $\hat a^{2}$ and $\hat a^{\dagger 2}$ as well as constant shifts.
	
	If the resonator is driven coherently at frequency $\omega_{d}$, the intracavity field acquires a large coherent amplitude. Linearizing around the steady-state amplitude, $\hat a\rightarrow \alpha+\hat a$ (and similarly for $\hat a^\dagger$), and moving to the frame rotating at $\omega_d$, the Hamiltonian can be written as
	\begin{equation}
		H_{\mathrm{sys}}=\hbar\omega_{\mathrm{m}}\hat{b}^{\dagger}\hat{b}+\hbar\omega_{\Delta}\hat{a}^{\dagger}\hat{a}
		+\hbar G\,(\hat{a}+\hat{a}^{\dagger})(\hat{b}+\hat{b}^{\dagger})+\cdots,
		\label{SA1E04}%
	\end{equation}
	where $\omega_{\Delta}=\omega_{\mathrm{c}}-\omega_{d}$ is the detuning,
	$G=\sqrt{\bar{n}}G_{0}$ is the linearized coupling, $\bar{n}=\left\vert
	\alpha\right\vert ^{2}$ is the incavity photon number, and the ellipsis
	denotes higher-order fluctuation terms that are neglected in the standard
	linearized treatment.

	Including dissipation and input noise, one obtains quantum Langevin equations for the mechanical quadratures and resonator quadratures, together with the input--output formula (for example, for the measured output quadrature),
	\begin{equation}
		q_{\mathrm{out}}(t)=q_{\mathrm{in}}(t)+\sqrt{\kappa}\,q(t),
	\end{equation}
	where $\kappa$ is the resonator linewidth. These relations provide the basis for the force/force-gradient readout analysis in the main text.
	
	\section{Displacement-to-flux transduction}
	
	As illustrated in Fig.~1 of the main text, a displacement $x$ of the ferromagnetic microsphere changes the flux $\Phi_{P}$ threading the SQUID pickup loop. The readout circuit experiences a corresponding flux
	\begin{equation}
		\Phi=\lambda_{\Phi}\Phi_{P},
	\end{equation}
	where $\lambda_{\Phi}$ is a conversion factor set by the mutual inductance between the pickup loop and the SQUID/readout circuit.
	
	For a pickup loop of radius $r_{\mathrm{s}}$ located a distance $l$ from the microsphere and oriented at a polar angle $\theta$ (with respect to the vertical axis passing through the sphere), the pickup flux is
	\begin{equation}
		\Phi_{P}=\iint_{S}\vec{B}(x)\cdot d\vec{S},
		\label{SA2E01}%
	\end{equation}
	where $\vec{B}(x)$ is the microsphere magnetic field at the loop. When the sphere radius is much smaller than the separation ($R\ll l$), the field can be accurately approximated by a magnetic dipole field with dipole moment $\mu$. In this limit, the displacement-to-flux coefficient becomes
	\begin{equation}
		\eta=\lambda_{\Phi}\left|\frac{\partial \Phi_{P}}{\partial x}\right|
		=\lambda_{\Phi}\beta\!\left(r_{\mathrm{s}}/l,\theta\right) \frac{\mu_{0}\mu\, r_{\mathrm{s}}^{2}}{l^{4}},
		\label{SA2E02}%
	\end{equation}
	where $\mu_{0}$ is the vacuum permeability and $\beta(r_{\mathrm{s}}/l,\theta)$ is a dimensionless geometric factor determined by the loop geometry and its relative placement.
	
	In the small-loop limit $r_{\mathrm{s}}\ll l$, $\beta(r_{\mathrm{s}}/l,\theta)$ is maximized at $\cos\theta=\sqrt{11/15}$ (i.e., $\theta\simeq 31^{\circ}$), giving $\beta_{\max}=8/\sqrt{15}\simeq 2.1$. Unless stated otherwise, the numerical estimates in the main text use representative values given in Tab.\ref{SA2E03}. 
	
		\begin{table}[h]
		\centering
		\caption{Representative parameters used in the numerical estimates.}
		\label{SA2E03}
		\begin{tabular*}{\linewidth}{@{\hspace{6pt}\extracolsep{\fill}}cccccc@{\hspace{6pt}}}
			\hline\hline
			$\lambda_{\Phi}$ & $\beta$ & $l$ & $\kappa$ & $\omega_{\mathrm{c}}$ & $(d\omega_{\mathrm{c}}/d\Phi)/(2\pi)$ \\
			\hline
			-- & -- & \rm{cm} & \rm{MHz} & \rm{GHz} & $\rm{Hz/Wb}$ \\
			\hline
			$0.1$ & $2.1$ & $2.0$ & $1.0$ & $10$ & $4.8\times10^{23}$ \\
			\hline\hline
		\end{tabular*}
	\end{table}
	
	\section{Non-Newtonian gravitational force gradient}
	
	A Yukawa-type modification to Newtonian gravity between two point masses $m_{1}$ and $m_{2}$ separated by $|r_{12}|$ can be written as
	\begin{equation}
		V(|r_{12}|)=-\frac{Gm_{1}m_{2}}{|r_{12}|}\left(1+\alpha e^{-|r_{12}|/\lambda}\right),
	\end{equation}
	where $G$ is the gravitational constant, $\alpha$ is the dimensionless coupling strength, and $\lambda$ is the interaction range. The Newtonian contribution from an infinite uniform plane produces a constant acceleration and therefore a \emph{zero} force gradient; it does not shift the oscillator frequency. Hence, for frequency-shift (force-gradient) sensing, the relevant contribution is the Yukawa term.
	
	Consider a point test mass element $dm$ at a height $l$ above an infinite plane of uniform density $\rho_{0}$. The Yukawa potential energy due to the plane is
	\begin{equation}
		dV=-\alpha G\,dm\int \frac{e^{-r/\lambda}}{r}\,dm_{0},
		\label{SA3E01}%
	\end{equation}
	where the integral is over the plane mass elements $dm_{0}$. Using spherical shells centered at the test mass, one has $dm_{0}=\rho_{0}\,dV_{0}=\rho_{0}\,\Omega\,r^{2}dr$ with solid angle $\Omega=2\pi(1-\cos\theta)=2\pi(1-l/r)$. Performing the integral yields
	\begin{equation}
		dV=-2\pi\alpha G\rho_{0}\lambda^{2}e^{-l/\lambda}\,dm.
		\label{SA3E02}%
	\end{equation}
	
	Now consider a ferromagnetic sphere of radius $R$ and density $\rho$ whose center is a distance $d$ above the plane. Slicing the sphere by horizontal planes at coordinate $z$ (measured from the center), the slice at $z$ has thickness $dz$, mass
	\begin{equation}
		dm=\rho\pi(R^{2}-z^{2})\,dz,
	\end{equation}
	and distance to the plane $l=R+d-z$. Integrating \eqref{SA3E02} over the sphere volume gives the Yukawa potential energy
	\begin{equation}
		V(d)=-4\pi^{2}\alpha G\rho_{0}\rho\,\lambda^{5}e^{-d/\lambda}\,C(R,\lambda),
		\label{SA3E03}%
	\end{equation}
	with geometric factor
	\begin{equation}
		C(R,\lambda)=e^{-2R/\lambda}\left(R/\lambda+1\right)+\left(R/\lambda-1\right).
	\end{equation}
	Differentiating with respect to $d$ gives the Yukawa force and the force gradient,
	\begin{align}
		F_{G}(d) &  =-\frac{\partial V(d)}{\partial d}=-4\pi^{2}\alpha\rho_{0}%
		\rho\lambda^{4}Ge^{-d/\lambda}C(R,\lambda),\label{SA3E04a}\\
		F_{G}^{\prime}(d) &  =\frac{\partial^{2}V(d)}{\partial d^{2}}=4\pi^{2}%
		\alpha\rho_{0}\rho\lambda^{3}Ge^{-d/\lambda}C(R,\lambda).\label{SA3E04b}%
	\end{align}
	Fitting experimental force-gradient data to \eqref{SA3E04b} provides constraints on $|\alpha|$ as a function of the interaction range $\lambda$.
	
	\section{Parameter estimation and chi-square analysis}
	
	Consider $N$ experimental data points $(x_{i},y_{i})$ and a model $y=f(x;p)$ with fit parameter(s) $p$ (additional parameters may be treated as fixed). Assume independent Gaussian measurement errors in $y$ with standard deviation $\Delta y_{i}$. The conditional probability density for each point is
	\begin{equation}
		p(y_{i}\,|\,x_{i},p)=\frac{1}{\sqrt{2\pi}\Delta y_{i}}\exp\!\left[
		-\frac{\left(y_{i}-f(x_{i};p)\right)^{2}}{2(\Delta y_{i})^{2}}\right].
		\label{eq:single_pdf}%
	\end{equation}
	Independence implies the likelihood
	\begin{equation}
		L(p)=\prod_{i=1}^{N}p(y_{i}\,|\,x_{i},p)=(2\pi)^{-N/2}\prod_{i}\frac
		{e^{-\chi^{2}(p)/2}}{\Delta y_{i}},\label{eq:like}%
	\end{equation}
	where the chi-square statistic is
	\begin{equation}
		\chi^{2}(p)=\sum_{i=1}^{N}\frac{\left[y_{i}-f(x_{i};p)\right]^{2}}{(\Delta y_{i})^{2}}.
		\label{eq04}%
	\end{equation}
	If the dependence of $\Delta y_{i}$ on $p$ can be neglected, maximizing $L(p)$ is equivalent to minimizing $\chi^{2}(p)$. With $\nu=N-m$ degrees of freedom ($m$ is the number of fit parameters), the reduced chi-square is
	\begin{equation}
		\chi_{\nu}^{2}(p)=\frac{\chi^{2}(p)}{\nu}.
		\label{eq:chi2red}%
	\end{equation}
	The best-fit parameter $p_{0}$ satisfies
	\begin{equation}
		p_{0}=\arg\min_{p}\,\chi_{\nu}^{2}(p).
		\label{eq:phat}%
	\end{equation}
	
	Near the minimum, a standard Gaussian approximation leads to the $\Delta\chi^{2}$ criterion for a $1\sigma$ confidence interval in a \emph{single} fit parameter:
	\begin{equation}
		\chi^{2}(p_{0}\pm \Delta p)-\chi^{2}_{\min}=1.
		\label{eq:dchi2red}%
	\end{equation}
	Expanding $\chi^{2}(p)$ to second order about $p_{0}$,
	\begin{equation}
		\chi^{2}(p)\simeq \chi_{\min}^{2}+\frac{1}{2}\left.\frac{d^{2}\chi^{2}}{dp^{2}}\right|_{p_{0}}(p-p_{0})^{2},
		\label{eq:taylor2}%
	\end{equation}
	one obtains an estimate of the parameter standard deviation,
	\begin{equation}
		\sigma_{p}\simeq\sqrt{2/\left\vert  \dfrac{d^{2}\chi^{2}}{dp^{2}}\right\vert
			_{p_{0}}}.\label{eq:sigmap}%
	\end{equation}
	Physically, the curvature quantifies how rapidly the goodness of fit degrades when the parameter is perturbed: larger curvature implies stronger statistical constraint.
	
	If the abscissa values $x_{i}$ have uncertainties $\Delta x_{i}$, one may incorporate them (when sufficiently small) by linear error propagation. Expanding
	\begin{equation}
		f(x_{i}+\delta x;p)\simeq f(x_{i};p)+f'(x_{i};p)\,\delta x,
		\label{eq:taylor}%
	\end{equation}
	with $f'(x_{i};p)=\left.\partial f/\partial x\right|_{x_{i}}$, and assuming the $x$ and $y$ noises are independent, the effective variance in $y$ is
	\begin{equation}
		(\Delta y_{i,\mathrm{eff}})^{2}\equiv (\Delta y_{i})^{2}+\left[f'(x_{i};p)\,\Delta x_{i}\right]^{2}.
		\label{eq:sigma_eff}%
	\end{equation}
	Replacing $\Delta y_{i}$ by $\Delta y_{i,\mathrm{eff}}$ in \eqref{eq04}--\eqref{eq:chi2red} yields a consistent weighted fit. When $f'(x_{i};p)$ depends appreciably on $p$, one typically evaluates $f'$ at $p=p_{0}$, or iterates the fit by updating $f'$ after obtaining an initial estimate.
	
	Under the ideal assumptions (correct model, correct error bars, independent points), $\chi^{2}$ follows a chi-square distribution with $\nu$ degrees of freedom, implying $\mathbb{E}[\chi_{\nu}^{2}]\simeq 1$. Thus:
	\begin{itemize}
		\item $\chi_{\nu,\min}^{2}\simeq 1$ indicates statistically self-consistent residuals and uncertainties.
		\item $\chi_{\nu,\min}^{2}\gg 1$ often indicates model inadequacy, unaccounted systematic errors, or underestimated uncertainties.
		\item $\chi_{\nu,\min}^{2}\ll 1$ often indicates overestimated uncertainties or correlations that reduce the effective degrees of freedom.
	\end{itemize}
	In summary, $\chi_{\nu,\min}^{2}$ quantifies global consistency, while the curvature of $\chi^{2}(p)$ near its minimum sets the statistical uncertainty of the fitted parameter $p_{0}$.

	%\nocite{1}
	\nocite{*}
	\bibliography{ref.bib}
	
\end{document}